\documentclass[pdflatex,sn-mathphys-num]{sn-jnl}% Math and Physical Sciences Numbered Reference Style 
\usepackage{graphicx}%
\usepackage{multirow}%
\usepackage{amsmath,amssymb,amsfonts}%
\usepackage{amsthm}%
\usepackage{mathrsfs}%
\usepackage[title]{appendix}%
\usepackage{xcolor}%
\usepackage{textcomp}%
\usepackage{manyfoot}%
\usepackage{booktabs}%
\usepackage{algorithm}%
\usepackage{algorithmicx}%
\usepackage{algpseudocode}%
\usepackage{listings}%
\usepackage{graphicx}       % graphics
\usepackage{subcaption}

\theoremstyle{thmstyleone}%
\theoremstyle{thmstyletwo}%

\theoremstyle{thmstylethree}%

\begin{document}

\title[Article Title]{DeepSilencer: A Novel Deep Learning Model for Predicting siRNA Knockdown Efficiency}


% \author*[1,2]{\fnm{First} \sur{Author}}\email{iauthor@gmail.com}



% \affil*[1]{\orgdiv{Department}, \orgname{Organization}, \orgaddress{\street{Street}, \city{City}, \postcode{100190}, \state{State}, \country{Country}}}

% \affil[2]{\orgdiv{Department}, \orgname{Organization}, \orgaddress{\street{Street}, \city{City}, \postcode{10587}, \state{State}, \country{Country}}}

% \affil[3]{\orgdiv{Department}, \orgname{Organization}, \orgaddress{\street{Street}, \city{City}, \postcode{610101}, \state{State}, \country{Country}}}

\author[1]{\fnm{Wangdan} \sur{Liao}}\email{liaowangdan@buaa.edu.cn}

\author*[1,2]{\fnm{Weidong} \sur{Wang}}\email{wangwd301@126.com}



\affil[1]{\orgdiv{School of Biological Science and Medical Engineering}, \orgname{Beihang University}, \orgaddress{\street{37 Xueyuan Road}, \city{Beijing}, \postcode{100191}, \country{China}}}

\affil*[2]{\orgdiv{Medical Innovation Research Division}, \orgname{Chinese PLA General Hospital}, \orgaddress{\street{28 Fuxing Road}, \city{Beijing}, \postcode{100853}, \country{China}}}


\abstract{

\textbf{Background:} Small interfering RNA (siRNA) is a promising therapeutic agent due to its ability to silence disease-related genes via RNA interference. While traditional machine learning and early deep learning methods have made progress in predicting siRNA efficacy, there remains significant room for improvement. Advanced deep learning techniques can enhance prediction accuracy, reducing the reliance on extensive wet-lab experiments and accelerating the identification of effective siRNA sequences. This approach also provides deeper insights into the mechanisms of siRNA efficacy, facilitating more targeted and efficient therapeutic strategies.

\textbf{Methods:} We introduce DeepSilencer, an innovative deep learning model designed to predict siRNA knockdown efficiency. DeepSilencer utilizes advanced neural network architectures to capture the complex features of siRNA sequences. Our key contributions include a specially designed deep learning model, an innovative online data sampling method, and an improved loss function tailored for siRNA prediction. These enhancements collectively boost the model's prediction accuracy and robustness.

\textbf{Results:} Extensive evaluations on multiple test sets demonstrate that DeepSilencer achieves state-of-the-art performance using only siRNA sequences and basic physicochemical properties. Our model surpasses several other methods and shows superior predictive performance, particularly when incorporating thermodynamic parameters.

\textbf{Conclusion:} The advancements in data sampling, model design, and loss function significantly enhance the predictive capabilities of DeepSilencer. These improvements underscore its potential to advance RNAi therapeutic design and development, offering a powerful tool for researchers and clinicians.}


\maketitle

\section{Background}\label{sec1}

RNA interference (RNAi) technology has emerged as a promising therapeutic approach by enabling the sequence-specific silencing of disease-related genes. Small interfering RNA (siRNA) is central to this process, being incorporated into the RNA-induced silencing complex (RISC), which includes proteins such as Argonaute-2 (Ago-2), Dicer, and TRBP\cite{zamore2000rnai}, \cite{tang2024rnai} and \cite{pratt2009rna}. The antisense strand of siRNA guides RISC to the target mRNA, where Ago-2 mediates cleavage, resulting in gene knockdown\cite{deerberg2013minimal}. This mechanism allows siRNA to target traditionally undruggable genes, significantly expanding potential therapeutic targets.

Designing highly efficient siRNAs remains challenging. Effective siRNAs must avoid triggering the innate immune system, achieve high specificity in target cleavage, and minimize off-target effects. Consequently, designing effective siRNA is crucial for RNAi therapeutics, prompting the development of various computational tools. Tools like OligoWalk, siRNAPred, and i-score have limitations due to dataset biases, prediction methods, and off-target effects\cite{lu2008oligowalk}, \cite{han2017utilizing} and \cite{ichihara2007thermodynamic}.

Early machine learning methods, such as LASSO regression, support vector machines (SVM), and linear regression, improved prediction accuracy but were limited by their simple structures, which failed to capture complex features\cite{lu2008efficient}, \cite{lu2008fundamental}. With the advent of deep learning, researchers began leveraging advanced models for better understanding siRNA sequences.

Deep learning brought significant improvements. CNNs and DNNs automatically extracted features from siRNA sequences, enhancing accuracy\cite{han2018sirna}. Graph Neural Networks (GNNs) modeled siRNA-mRNA interactions, incorporating thermodynamic parameters\cite{ijms232214211}. Transformer-based models have excelled in natural language processing and genomics due to their ability to handle sequential data through attention mechanisms\cite{10.5555/3295222.3295349}. Models like BERT-siRNA leverage pre-training on extensive human genomic sequences to enhance siRNA efficacy prediction\cite{devlin2018bert}, \cite{xu2024bert}. OligoFormer combines Transformers with LSTM and convolutional layers to capture deep sequence features and complex siRNA-mRNA interaction patterns\cite{hochreiter1997long}, \cite{lecun1989backpropagation} and \cite{bai2024oligoformer}.

Despite these advancements, existing models face limitations in interpretability, usability, robustness, and data utilization. In response, we propose DeepSilencer, a deep learning model specifically designed for predicting siRNA knockdown efficiency. Our contributions include:

\begin{enumerate}
\item \textbf{Innovative Model Design}: Combines transformer and CNN modules to effectively capture deep features of siRNA sequences, enhancing prediction accuracy and robustness without pre-training.
\item \textbf{Versatile Data Input}: Supports siRNA sequences alone or combined with thermodynamic features.
\item \textbf{Enhanced Loss Function}: Improved loss functions tailored for siRNA prediction tasks, boosting accuracy and robustness.
\item \textbf{Comprehensive Improvements}: Achieves state-of-the-art results using only siRNA sequences and simple physicochemical properties.
\end{enumerate}

In summary, our work builds on existing methods' strengths while addressing their limitations, offering a robust solution for siRNA knockdown efficiency prediction.

\section{Methods}\label{sec2}

We have implemented a series of innovative approaches to improve the accuracy and robustness of siRNA silencing efficiency prediction. Initially, in response to the limited quantity of siRNA data, we introduced a novel dynamic random data pairing mechanism, namely Selective Pair Sampling, to fully exploit the potential of the available data. Subsequently, given the challenge of aligning different publicly available datasets on siRNA silencing efficiency, we designed a novel multi-task system named Tri-Task Learning Framework. To further refine this framework, we crafted an improved loss function that significantly boosts the model's accuracy and robustness. Finally, we designed DeepSilencer, a model specifically tailored for RNA sequence analysis. By integrating Transformer and Convolutional Neural Network (CNN) architectures, DeepSilencer effectively captures the intricate features of siRNA sequences, thereby improving prediction accuracy and robustness without the need for pre-training. Detailed introductions of these components will be presented in the following sections.

\subsection{Data Processing}

We utilized the datasets summarized by OligoFormer, which aggregated nine studies, resulting in a total of 3,714 siRNAs and 75 mRNAs\cite{huesken2005design}, \cite{katoh2007specific}, \cite{amarzguioui2003tolerance}, \cite{harborth2003sequence},\cite{reynolds2004rational}, \cite{vickers2003efficient}, \cite{ui2004guidelines} and \cite{hsieh2004library}. These datasets were organized into three categories: the Huesken dataset, the Takayuki dataset, and a mixed set (Mixset) comprising the remaining data. The inhibition efficacy of siRNAs was normalized to a range of 0 to 100\%, with a 70\% inhibition threshold distinguishing positive from negative siRNAs.

To ensure consistency across datasets, all siRNA sequences were truncated to 19 nucleotides. This standardization facilitated uniform processing and feature extraction. OligoFormer extracted thermodynamic parameters for each siRNA, including stability and binding affinity, essential for predicting siRNA efficacy. Detailed thermodynamic parameters are provided in Table \ref{table:tr}. Subsequently, considering the significance of these parameters, we conducted additional analyses on them. Recognizing the complexity of feature interactions, we employed a random forest model to fit the silencing efficiency solely based on thermodynamic features and obtained feature importance scores. Although the feature importance analysis based on the random forest model indicated that the importance levels of some features were relatively low, we discovered that removing these features would affect the performance of our model. Therefore, after careful consideration, we ultimately decided to retain these features to maintain the overall performance and stability of the model.

Our model, DeepSilencer, employs these processed datasets for training and validation without requiring complex pre-training. It supports two prediction modes: one using only siRNA sequence data and the other using both siRNA sequences and thermodynamic parameters. In subsequent experiments, we will evaluate the model's performance under both conditions.

\begin{table}[ht]
\centering
\caption{Thermodynamic Parameters for siRNA-mRNA Binding}
\begin{tabular}{|c|l|c|}
\hline
\textbf{Parameter} & \textbf{Description} & \textbf{Feature Contribution} \\ \hline
\(\Delta\Delta G_{\text{ends}}^\circ\) & End free energy difference & 0.264\\ \hline
\(\Delta G^\circ_1\) & Free energy change of the first base pair & 0.039\\ \hline
\(\Delta H^\circ_1\) & Enthalpy change of the first base pair & 0.018\\ \hline
\(U_1\) & Uracil content of the first base pair & 0.023\\ \hline
\(G_1\) & Guanine content of the first base pair & 0.003\\ \hline
\(\Delta H^\circ_{\text{all}}\) & Total enthalpy change of all base pairs & 0.218\\ \hline
\(U_{\text{all}}\) & Total uracil content of all base pairs & 0.052\\ \hline
\(UU_1\) & Uracil content of the first base pair and its neighbors & 0.003\\ \hline
\(G_{\text{all}}\) & Total guanine content of all base pairs & 0.054\\ \hline
\(GG_1\) & Guanine content of the first base pair and its neighbors & 0.002\\ \hline
\(GC_1\) & Guanine and cytosine content of the first base pair & 0.001\\ \hline
\(GG_{\text{all}}\) & Total guanine content of all base pairs & 0.033\\ \hline
\(\Delta G^\circ_2\) & Free energy change of the second base pair & 0.046\\ \hline
\(UA_{\text{all}}\) & Total uracil and adenine content of all base pairs & 0.022\\ \hline
\(U_2\) & Uracil content of the second base pair & 0.007\\ \hline
\(C_1\) & Cytosine content of the first base pair & 0.004\\ \hline
\(CC_{\text{all}}\) & Total cytosine content of all base pairs & 0.033\\ \hline
\(\Delta G^\circ_{18}\) & Free energy change of the eighteenth base pair & 0.037\\ \hline
\(CC_1\) & Cytosine content of the first base pair and its neighbors & 0.004\\ \hline
\(GC_{\text{all}}\) & Total guanine and cytosine content of all base pairs & 0.031\\ \hline
\(CG_1\) & Cytosine and guanine content of the first base pair & 0.007\\ \hline
\(\Delta G^\circ_{13}\) & Free energy change of the thirteenth base pair & 0.051\\ \hline
\(UU_{\text{all}}\) & Total uracil content of all base pairs & 0.028\\ \hline
\(A_{19}\) & Adenine content of the nineteenth base pair & 0.013\\ \hline
\end{tabular}
\label{table:tr}
\end{table}

\subsection{Selective Pair Sampling}

Due to the limited availability of siRNA data, with the largest dataset containing fewer than 3000 siRNAs, it is crucial to maximize the potential of the available data and enhance the model's ranking capability. We designed a novel pair loss-based approach for siRNA silencing efficiency prediction\cite{NIPS2016_6b180037}. By transforming each training sample from a single siRNA to a randomly paired siRNA pair, this method not only increases data diversity but also improves the model's ranking ability through pair loss.

To achieve this, we implemented an innovative dynamic random data pairing mechanism to sample siRNA pairs that meet specific criteria (see 
 Algorithm \ref{alg:selective_pair_sampling}). This method, called \textbf{Selective Pair Sampling}, ensures that the sampled siRNA pairs have label differences within a predefined range, thereby enhancing the model's ability to learn from subtle differences in siRNA efficacy.

The \textbf{Selective Pair Sampling} method operates as follows:

\begin{enumerate}
    \item \textbf{Random Initial Selection}: For each data retrieval operation, an initial siRNA sequence (\texttt{$s_1$}) with a length of 19 nucleotides and its corresponding label (\texttt{$y_1$}) is selected from the dataset.
    \item \textbf{Conditional Pairing}: The method then enters a loop where it randomly selects a second siRNA sequence (\texttt{$s_2$}) and its label (\texttt{$y_2$}). This loop continues until a pair is found where the absolute difference between their labels (\texttt{$y_1$} and \texttt{$y_2$}) falls within the range of $\alpha_1$ to $\alpha_2$. This condition ensures that the selected pairs exhibit a moderate difference in inhibition efficiency, which is crucial for training models to discern subtle variations.
    \item \textbf{Data Preparation}: Once a suitable pair is identified, both sequences are tokenized using a tokenizer. The tokenized sequences are padded and truncated to a maximum length of 21 nucleotides, with a \texttt{[CLS]} token added at the beginning and a \texttt{[SEQ]} token at the end.
    \item \textbf{Return Values}: The method returns the tokenized sequences, their corresponding normalized silencing efficiencies, thermodynamic parameters(TR), and their class labels.
\end{enumerate}

\begin{algorithm}
\caption{Selective Pair Sampling}\label{alg:selective_pair_sampling}
\begin{algorithmic}[1]
\State \textbf{Input:} Dataset $D$ with siRNA sequences ($s$), thermodynamic parameters ($tr$), and labels ($y$)
\State \textbf{Output:} Paired siRNA sequences and features

\For{each data retrieval operation}
    \State $s_1, tr_1, y_1 \gets$ Randomly select an siRNA sequence and its label from $D$
    \While{True}
        \State $s_2, tr_2, y_2 \gets$ Randomly select another siRNA sequence and its label from $D$
        \If{$\alpha_1 < |y_1 - y_2| < \alpha_2$}
            \State \textbf{break}
        \EndIf
    \EndWhile
    \State Sample the pair $(s_1, s_2)$, their thermodynamic parameters $(tr_1, tr_2)$, and their respective labels $(y_1, y_2)$
\EndFor

\end{algorithmic}
\end{algorithm}

\subsection{Multi-Task Learning for siRNA Silencing Efficiency Prediction}

Existing publicly available data on siRNA silencing efficiency are derived from various experimental measurements. However, due to differences in experimental methodologies and measurement techniques, aligning the values across different datasets can be challenging. In the OligoFormer study, the authors addressed this issue by converting continuous values into categorical labels. Specifically, the inhibition efficacy or activity of siRNAs in all datasets was normalized to a range from 0 to 100\%, and a threshold of 70\% maximum inhibition was used to classify siRNAs as positive or negative.

However, this approach introduces a problem: siRNAs with similar efficacy near the classification boundary may end up with distinctly different labels. To address this, we designed a novel multi-task system, named \textbf{Tri-Task Learning Framework}, which consists of three tasks\cite{zhang2018overview}:

\begin{enumerate}
    \item \textbf{Classification Task}: Similar to OligoFormer, this task classifies siRNAs based on a predefined threshold.
    \item \textbf{Regression Task}: This task directly predicts the normalized silencing efficiency.
    \item \textbf{Contrastive Task}: This task predicts the difference in silencing efficiency between pairs of siRNAs, aiming to improve the model's ranking capability.
\end{enumerate}

The total loss for the Tri-Task Learning Framework is a combination of the losses from the three tasks:

\begin{itemize}
    \item \textbf{Classification Loss} (\(\mathcal{L}_{\text{cls}}\)): This is typically a binary cross-entropy loss for classifying siRNAs as positive or negative. This loss is specifically for $\rm{siRNA_1}$ and does not include $\rm{siRNA_2}$.
    \item \textbf{Regression Loss} (\(\mathcal{L}_{\text{reg}}\)): This is typically a Smooth L1 Loss (also known as Huber Loss) for predicting the normalized silencing efficiency. This loss includes both $\rm{siRNA_1}$ and $\rm{siRNA_2}$.
    \item \textbf{Contrastive Loss} (\(\mathcal{L}_{\text{cont}}\)): This loss measures the ability of the model to predict the difference in silencing efficiency between pairs of siRNAs. One common approach is to use a margin-based contrastive loss. This loss has a coefficient of 2.
\end{itemize}

The Smooth L1 Loss, also known as Huber Loss, is defined as:

\[
\mathcal{L}_{\text{smooth\_L1}}(x, y) = 
\begin{cases} 
\frac{1}{2} \frac{(x - y)^2}{\beta} & \text{if } |x - y| < \beta \\
|x - y| - \frac{1}{2} \beta & \text{otherwise}
\end{cases}
\]

Here, \(\beta\) is a hyperparameter that determines the point at which the loss function transitions from quadratic to linear. In our experiments, we set \(\beta = 2.4\)\cite{7485869}.

This loss function is less sensitive to outliers compared to the Mean Squared Error (MSE) loss, making it more robust for regression tasks where outliers might be present. It combines the best properties of both L1 and L2 losses by being quadratic when the error is small and linear when the error is large.

Given two siRNA sequences, \(\text{siRNA}_1\) and \(\text{siRNA}_2\), the model, DeepSilencer, predicts their respective silencing efficiencies, \( \hat{y}_1 \) and \( \hat{y}_2 \). The output of the contrastive model is defined as the difference between these predictions:

\[
\Delta \hat{y} = \text{DeepSilencer}(\text{siRNA}_1) - \text{DeepSilencer}(\text{siRNA}_2)
\]

where \(\Delta \hat{y}\) represents the predicted difference in silencing efficiency between the two siRNAs. 

The true difference in silencing efficiencies for the siRNA pair is denoted as \(\Delta y = y_1 - y_2\), where \(y_1\) and \(y_2\) are the actual silencing efficiencies of \(\text{siRNA}_1\) and \(\text{siRNA}_2\), respectively.

The contrastive loss is then computed using the Smooth L1 loss function, which is defined as:

\[
\mathcal{L}_{\text{cont}} = \mathcal{L}_{\text{smooth\_L1}}(\Delta \hat{y}, \Delta y)
\]

This formulation allows the model to effectively learn to predict the relative differences in silencing efficiency between pairs of siRNAs, which is crucial for its overall performance.

The total loss \(\mathcal{L}_{\text{total}}\) can be expressed as:

\[
\mathcal{L}_{\text{total}} = \mathcal{L}_{\text{cls}} + \mathcal{L}_{\text{reg}} + 2 \times \mathcal{L}_{\text{cont}}
\]

Here, the coefficient 2 in front of \(\mathcal{L}_{\text{cont}}\) is manually tuned based on our validation experiments. The initial motivation was to enhance the model's ranking capability.

\subsection{Model Architecture}

\begin{figure}[ht]
  \centering
  \includegraphics[width=0.5\textwidth]{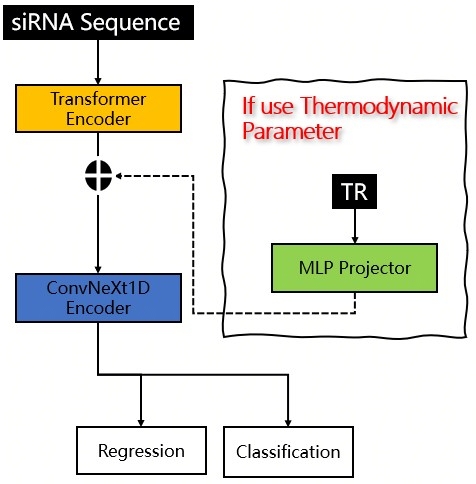}
  \caption{\textbf{DeepSilencer structure.} DeepSilencer is primarily composed of a Transformer encoder, consisting of multiple layers of transformers, and a ConvNeXt1D Encoder, formed by stacking multiple ConvNeXtBlocks in series\cite{woo2023convnext}. If thermodynamic parameters are included, a projection layer made of multilayer perceptrons(MLP) is used to transform the thermodynamic parameters into the hidden layer's dimensions. This transformation acts as a bias, added to the hidden vectors of each token obtained from the Transformer. Afterwards, the ConvNeXt Encoder performs dimensionality reduction on both the sequence and hidden layer dimensions of the sequence's hidden vectors. These vectors are then flattened and passed through two task-specific layers, each composed of MLP, to predict both the normalized knockdown efficiency value and the knockdown efficiency classification. Notably, the predicted efficiency value is related to both regression and contrastive tasks.
.}
  \label{fig:model_structure}
\end{figure}

Our model, \textbf{DeepSilencer}, integrates both Transformer and Convolutional Neural Network (CNN) architectures, specifically designed for RNA sequence analysis. The structure of the model is shown in Fig \ref{fig:model_structure}. The Transformer, through its self-attention mechanism, excels at capturing global dependencies within sequences, demonstrating strong modeling capabilities in both natural language processing and bioinformatics. However, siRNA sequences are typically short, and the available data is limited. Additionally, RNA molecular interactions frequently exhibit highly localized characteristics, making it challenging for a Transformer alone to sufficiently capture these local dependencies.

For instance, the self-attention mechanism of the Transformer performs exceptionally well when handling long sequences because it can simultaneously attend to all positions within the sequence at each layer. However, for short sequences like siRNA, an over-reliance on global information may lead to the model overlooking critical local features. Moreover, Transformer models generally require large datasets to effectively learn the complex relationships between sequences, and the limited size of siRNA datasets might hinder the model's generalization capabilities.

In contrast, Convolutional Neural Networks (CNNs), with their local receptive fields and weight-sharing mechanisms, are adept at extracting local features and possess a strong inductive bias. Studies have shown that CNNs exhibit robust performance when dealing with small-scale datasets. For example, CoAtNet combines convolutional and Transformer modules to achieve significant performance improvements on large-scale image datasets like ImageNet\cite{dai2021coatnet}. Similarly, Astroformer demonstrates that combining convolutional and Transformer structures and training from scratch can achieve excellent results on small-scale datasets\cite{Dagli2023AstroformerMD}.

Building on this foundation, we designed the \textbf{DeepSilencer} network, which leverages the strengths of both convolutional and Transformer architectures. The overall structure is as follows:

The first part of the model consists of a Transformer, which integrates global sequence information from siRNA. We utilize the T5 Encoder as the foundational structure, adjusting its configuration parameters as needed\cite{dai2021coatnet}. The T5 Encoder is a Transformer model that uses relative position encoding, here, we only utilized its architecture. Specifically, we configured a hidden layer dimension of 128, with 4 Transformer encoder layers, 4 attention heads, and a feed-forward network dimension of $128 \times 4$.

The latter part of the model comprises several one-dimensional ConvNeXtBlocks, which further enhance the integration of local information. ConvNeXt is an enhanced convolutional neural network architecture that improves performance through various innovations. ConvNeXtBlocks primarily consist of downsampling layers, depth-wise convolutions, LayerNorm, multilayer perceptrons (MLPs, equivalent to Feed Forward Layers in this context), and residual connections. In our implementation, we made a slight modification by substituting the original 2D convolutions with 1D convolutions. We incorporated multiple ConvNeXtBlocks to progressively extract and integrate local features:
\begin{itemize}
    \item The convolutional operations are divided into three stages, with downsampling achieved by setting the convolutional stride to 2 at the end of the first two stages.
    \item The first convolutional stage reduces the hidden layer dimension from 128 to 64, followed by a stride-2 convolutional layer for further downsampling. This stage contains 2 ConvNeXtBlocks.
    \item The second convolutional stage reduces the hidden layer dimension from 64 to 32, employing multiple stride-1 and stride-2 convolutional layers to gradually extract features. This stage comprises 3 ConvNeXtBlocks.
    \item The final convolutional stage maintains the dimension at 32 and includes 1 ConvNeXtBlock.
\end{itemize}
After the convolutional stages, a flattening operation merges the hidden and sequence dimensions before passing the features to the task-specific layers.

Finally, the model includes two task-specific fully connected layers for predicting siRNA knockdown efficiency and classification:
\begin{itemize}
    \item \textbf{Knockdown Efficiency Prediction Layer}: This layer consists of a dropout layer, a linear layer, and a GELU activation function, ultimately outputting a scalar value.
    \item \textbf{Knockdown Efficiency Classification Layer}: This layer, similar in structure, outputs the probabilities for two classes.
\end{itemize}

\begin{figure}[H]
  \centering
  % 第一行
  \begin{subfigure}[t]{0.4\textwidth}
    \vtop{\null\hbox{\includegraphics[width=\textwidth]{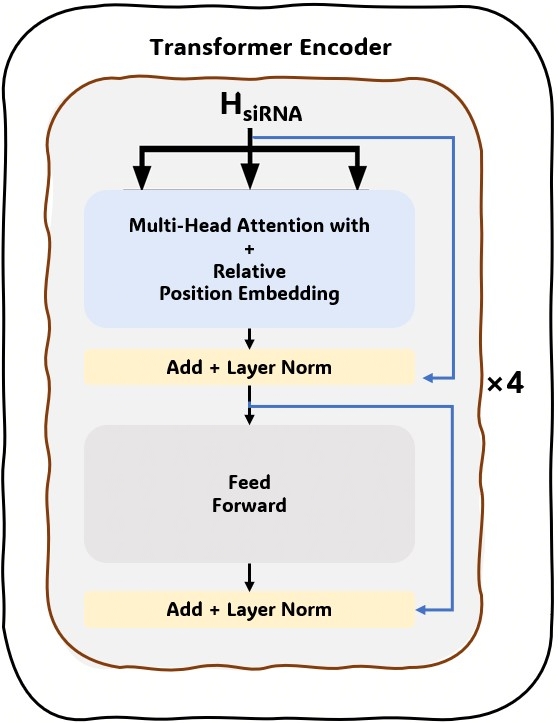}}}
    \caption{}
    \label{fig:transformer_encoder}
  \end{subfigure}
  \hspace{5mm}
  % \hfill % 添加空白或者 \hspace{5mm} 来调整子图间的水平间距
  \begin{subfigure}[t]{0.4\textwidth}
    \vtop{\null\hbox{\includegraphics[width=\textwidth]{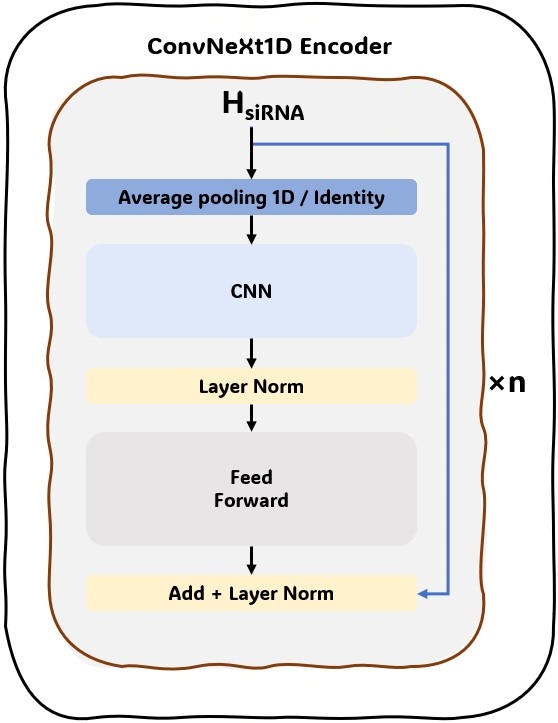}}}
    \caption{}
    \label{fig:convext_encoder}
  \end{subfigure}

\caption{(a) Transformer Encoder: This part of the model integrates global sequence information from siRNA using the T5 Encoder architecture. The configuration includes a hidden layer dimension of 128, 4 Transformer encoder layers, 4 attention heads, and a feed-forward network dimension of 128 × 4.
(b) ConvNeXt Encoder: This part of the model enhances the integration of local sequence information using several one-dimensional ConvNeXtBlocks. The ConvNeXt Encoder are divided into three stages: the first stage reduces the hidden layer dimension from 128 to 64 and includes 2 ConvNeXtBlocks; the second stage reduces the dimension from 64 to 32 with 3 ConvNeXtBlocks; the final stage maintains the dimension at 32 with 1 ConvNeXtBlock. Downsampling is achieved by setting the convolutional stride to 2 at the end of the first two stages.
}
  \label{fig:twu_structure}
\end{figure}

In summary, the \textbf{DeepSilencer} network effectively integrates global and local information, demonstrating superior performance in siRNA knockdown efficiency prediction and classification tasks.

\section{Results}\label{sec3}

\subsection{Evaluation on Within-Dataset Cross-Validation}

\begin{figure}
  \centering
  \includegraphics[width=0.8\textwidth]{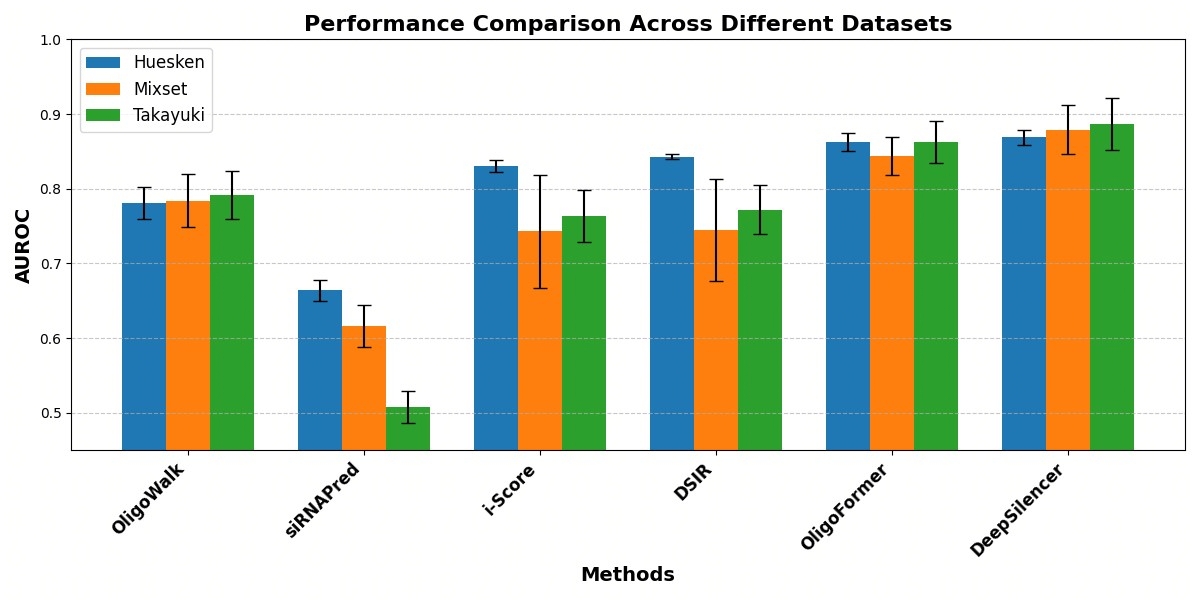}
  \caption{This figure illustrates the performance of various models evaluated using five-fold cross-validation across three datasets: Huesken, Mixset, and Takayuki. Each bar represents the mean Area Under the Receiver Operating Characteristic Curve for a specific model, with error bars indicating the standard deviation. The models compared include OligoWalk, siRNAPred, i-Score, DSIR, OligoFormer, and DeepSilencer. Notably, DeepSilencer demonstrates superior performance across all datasets, achieving the highest AUROC scores, as evident from the consistent pattern of the bars. Detailed numerical results, including mean values and standard deviations, can be found in Table \ref{table:performance_within}.}
  \label{fig:performance_within}
\end{figure}

To rigorously evaluate the performance of various models, we conducted five-fold cross-validation on three different datasets: Huesken, Mixset, and Takayuki. For each dataset, we calculated the mean and standard deviation of the Area Under the Receiver Operating Characteristic Curve (AUROC) for each model. The results are summarized in Fig.~\ref{fig:performance_within}. For detailed numerical results, please refer to Table~\ref{table:performance_within}.

\begin{table}[h]
\caption{AUROC on Within-Dataset Cross-Validation}\label{table:performance_within}%
\begin{tabular}{@{}llll@{}}
\toprule
Methods & Huesken  & Mixset & Takayuki\\
\midrule
OligoWalk    & 0.781 ± 0.021  & 0.7842 ± 0.035 & 0.792 ± 0.032 \\
siRNAPred    & 0.664 ± 0.014  & 0.6158 ± 0.028 & 0.508 ± 0.021 \\
i-Score      & 0.831 ± 0.008  & 0.743 ± 0.076 & 0.763 ± 0.035 \\
DSIR         & 0.843 ± 0.003  & 0.745 ± 0.068 & 0.772 ± 0.033 \\
OligoFormer  & 0.862 ± 0.012 & 0.844 ± 0.025 & 0.863 ± 0.028\\
DeepSilencer & \textbf{0.869 ± 0.010}  & \textbf{0.879 ± 0.033} & \textbf{0.887 ± 0.035} \\
\botrule
\end{tabular}
\end{table}

The results indicate that DeepSilencer consistently outperformed other models across all datasets. Notably, DeepSilencer demonstrated superior performance on the Mixset and Takayuki datasets, which are relatively smaller in size. This can be attributed to the model's simple and compact architecture, as well as its use of multi-task supervision, which enhances its ability to generalize well on smaller datasets.

\subsection{Evaluation on Cross-Dataset Validation
}

To evaluate the robustness and generalizability of our model, we conducted cross-dataset validation. Specifically, we trained our model on the Huesken dataset and tested it on a mixed set consisting of siRNA sequences from the Amarzguioui, Haborth, Hsieh, Khvorova, Reynolds, Vickers, and Ui-Tei datasets. The sequences in the mixed set are different from those in the training set, ensuring that the evaluation provides a stringent test of the model's robustness. We compared the performance of various publicly available models against our DeepSilencer model using the Area Under the Curve (AUC) and F1 score metrics.

As shown in Table~\ref{table:performance_inter}, our DeepSilencer model demonstrated superior performance across both metrics. Notably, even when using only the siRNA sequence information, DeepSilencer outperformed the previous state-of-the-art (SOTA) model, OligoFormer(siRNA). Furthermore, when incorporating siRNA sequence and thermodynamic features, DeepSilencer surpassed the performance of OligoFormer, which utilizes additional information such as siRNA, mRNA, RNAFM, and TR.

These results underscore the efficacy of our model in accurately predicting siRNA silencing efficiency, highlighting its potential for broader applications in RNA interference research.

\begin{table}[h]
\caption{Evaluation on Cross-Dataset Validation. The fields in parentheses represent the features used by the model: siRNA for siRNA sequences, mRNA for mRNA sequences, RNAFM for siRNA sequence embeddings extracted using the RNAFM model, and TR for thermodynamic parameters}\label{table:performance_inter}%
\begin{tabular}{@{}lll@{}}
\toprule
Methods & AUC  & F1 score\\
\midrule
OligoWalk    & 0.708  & 0.6258 \\
siRNAPred    & 0.578 & 0.525 \\
i-Score      & 0.746 & 0.219 \\
DSIR         & 0.748 & 0.693 \\
OligoFormer(siRNA+mRNA+RNAFM+TR)  & 0.815 & 0.769 \\
OligoFormer(siRNA)  & 0.787 & 0.742 \\
OligoFormer(siRNA+TR) & 0.793 & 0.751 \\
DeepSilencer(siRNA) & 0.802 & 0.760 \\
DeepSilencer(siRNA+TR) & \textbf{0.820} & \textbf{0.775} \\
\botrule
\end{tabular}
\end{table}

\subsection{Ablation Studies}

This section presents three ablation studies: \textit{Ablation Study on Various Data Sampling Methods}, \textit{Ablation Study on Different Types of Tasks}, and \textit{Ablation Study on Different Model Architectures}.We use AUROC to evaluate the performance throughout these studies.All the results are evaluated on cross-dataset.

\subsubsection*{Ablation Study on Various Data Sampling Methods}

For this study, we evaluated several data sampling methods: standard single-sample training, random pair sampling, Selective Pair Sampling (Easy), Selective Pair Sampling (Hard), and our default method, Selective Pair Sampling. The results are illustrated in Fig. \ref{fig:data_ablation}.

Selective Pair Sampling is defined by adjusting the threshold range parameters \(\alpha_1\) and \(\alpha_2\). For Selective Pair Sampling (Easy), we set \(\alpha_1 = 0.2\) and \(\alpha_2 = 0.4\), while for Selective Pair Sampling (Hard), we set \(\alpha_1 = 0.05\) and \(\alpha_2 = 0.1\). In the DeepSilencer model, the default values are \(\alpha_1 = 0.05\) and \(\alpha_2 = 0.2\). The results of more parameter combinations are presented in Table \ref{table:alpha_combination}. 

Our findings indicate that standard single-sample training (no pair sampling) performed the worst. The performance hierarchy observed was as follows: Selective Pair Sampling (Easy) $<$ random pair sampling $<$ Selective Pair Sampling (Hard), with the adjusted, moderate-difficulty Selective Pair Sampling achieving the best results. This aligns with previous research, which suggests that samples that are too easy or too hard can degrade model performance. Moderately difficult samples provide better gradients, thereby enhancing model learning.

\begin{table}[h]
\caption{The impact of different combinations of parameter $\alpha_1$ and $\alpha_2$ on the model.}\label{table:alpha_combination}%
\begin{tabular}{@{}lll@{}}
\toprule
$\alpha_1$ ~~~~~~~~~~~~~~~~~~~~~& $\alpha_2$  ~~~~~~~~~~~~~~~~~~~~~& AUROC\\
\midrule
0.2   & 0.4  & 0.808 \\
0.05  & 0.1 & 0.815 \\
0.05  & 0.2 & 0.820 \\
0.3   & 0.5 & 0.789 \\
0.2   & 0.5 & 0.804 \\
0.02  & 0.07 & 0.812 \\
\botrule
\end{tabular}
\end{table}

\subsubsection*{Ablation Study on Different Types of Tasks}

In this study, we compared regression tasks, classification tasks, contrastive tasks, and their combinations. The results are shown in Fig. \ref{fig:task_ablation}. 

For single tasks, the contrastive task yielded the best model performance. However, when combined into dual tasks, the combination of regression and classification tasks without the contrastive task performed the best. For tri-tasks, we evaluated using the predictions from the regression task and the classification task to calculate the AUROC. Ultimately, we found that using the regression task's predictions resulted in better performance.

It is evident that using only the classification task resulted in the poorest performance, which was comparable to the level of OligoWalk. However, simply switching from the classification task to the regression task improved the AUROC by 0.15. We attribute this to the fact that knockdown efficiency is inherently a continuous variable rather than a discrete one. For instance, the siRNA sequence \textbf{UCCAAAUCAAAGUUAUUUU} has a knockdown efficiency of \textbf{0.5212} with a classification label of \textbf{0}, while the sequence \textbf{AUUCUUCAGCUAGGUCAGC} has a knockdown efficiency of \textbf{0.5287} with a classification label of \textbf{1}. These two siRNAs should be considered equivalent in terms of their effect, but using classification labels forces a separation between them, which clearly impairs the model's ability to learn effectively.

\subsubsection*{Ablation Study on Different Model Architectures}

\begin{figure}[H]
  \centering
  % 第一行
  \begin{subfigure}[t]{0.47\textwidth}
    \vtop{\null\hbox{\includegraphics[width=\textwidth]{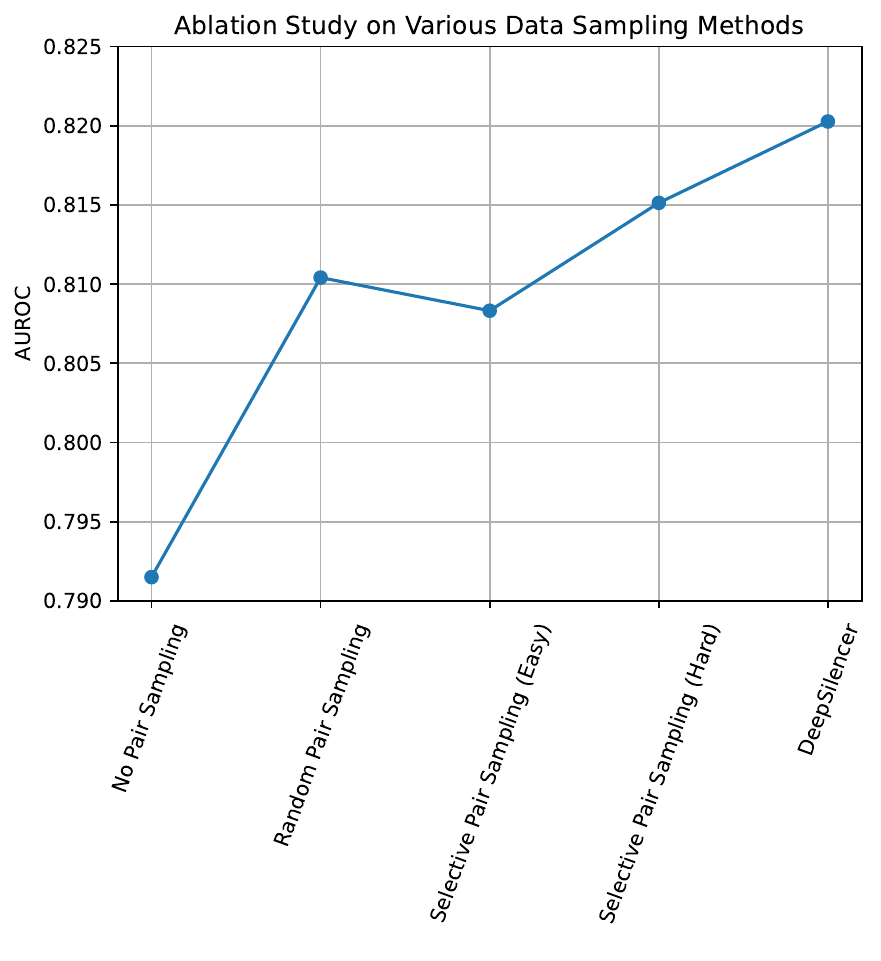}}}
    \caption{}
    \label{fig:data_ablation}
  \end{subfigure}
  % \hfill % 添加空白或者 \hspace{5mm} 来调整子图间的水平间距
  \begin{subfigure}[t]{0.47\textwidth}
    \vtop{\null\hbox{\includegraphics[width=\textwidth]{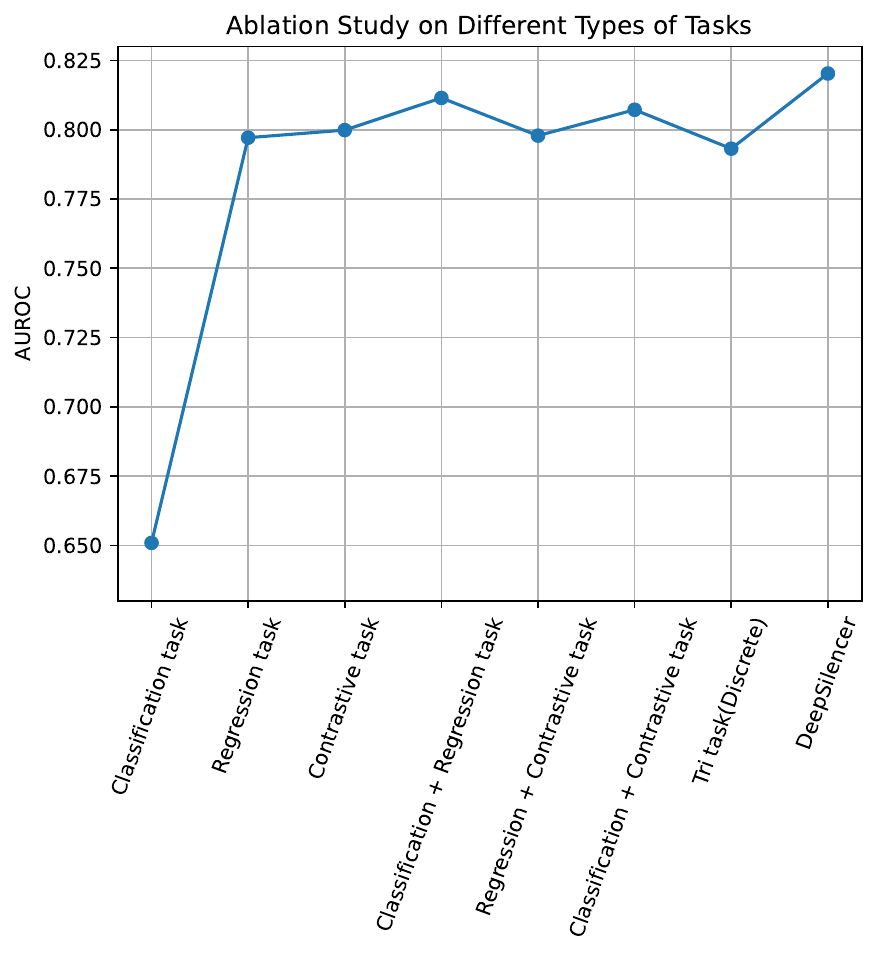}}}
    \caption{}
    \label{fig:task_ablation}
  \end{subfigure}
  
  % 第二行
  \begin{subfigure}[t]{0.94\textwidth}
    \centering
    \includegraphics[width=\textwidth]{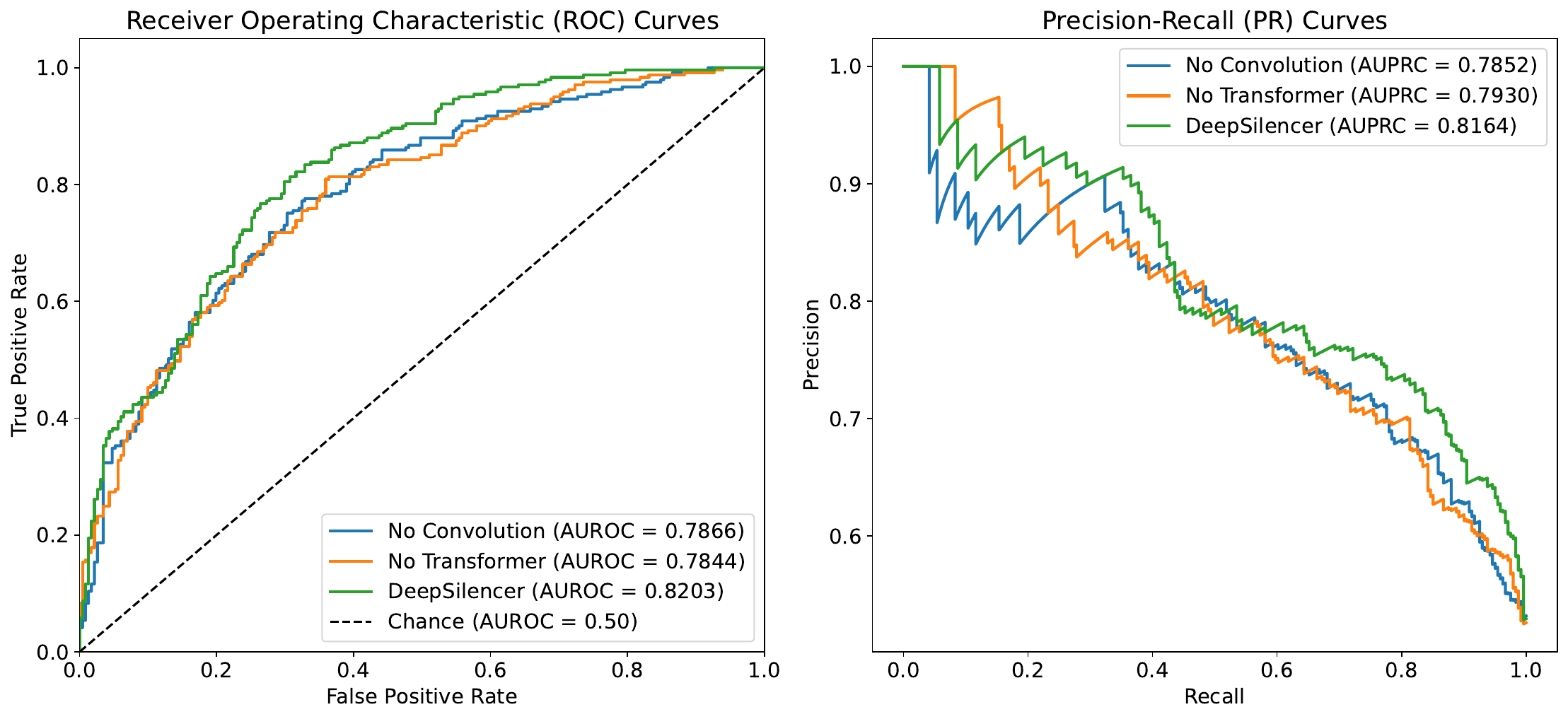}
    \caption{}
    \label{fig:model_ablation}
  \end{subfigure}

\caption{Ablation study. (a) Impact of different data sampling methods on model performance. "No pair sampling" involves random sample selection without contrastive loss. "Random pair sampling" selects two siRNA samples randomly for contrastive loss. "Selective Pair Sampling" adjusts \(\alpha_1\) and \(\alpha_2\) to control knockdown efficiency differences, making comparisons more challenging. "DeepSilencer" represents our optimal settings. (b) Impact of different tasks on model performance. We evaluated three single tasks (regression, classification, contrastive), three dual-task combinations, and two tri-task settings. "Tri-task (Discrete)" uses classification predictions, while "DeepSilencer" uses regression predictions, showing better performance. (c) Results of different model architectures: using only Transformer, only ConvNeXt (with an embedding layer), and the full DeepSilencer model. Evaluations based on AUROC and AUPRC show that DeepSilencer achieved the best performance.}

  \label{fig:ablation}
\end{figure}

The final ablation study focused on different model architectures. We computed both the AUROC and the Area Under the Precision-Recall Curve (AUPRC) to evaluate performance. The results, depicted in Fig. \ref{fig:model_ablation}, show that our hybrid model, combining transformer and CNN components, outperformed the others.

These ablation studies provide valuable insights into the optimal configurations for data sampling methods, task types, and model architectures, contributing to the overall improvement of our model's performance.

\section{Discussion}\label{sec4}

In this study, we introduce DeepSilencer, a model developed to predict the knockdown efficiency of siRNA. A key innovation of DeepSilencer lies in combining Selective Pair Sampling with a multi-task learning framework that incorporates contrastive loss. This combination enhances model performance by effectively distinguishing between siRNA sequences with similar knockdown efficiencies and is particularly beneficial in smaller datasets. Our approach harmoniously integrates different tasks, resulting in significant model improvements (see Fig. \ref{fig:task_ablation}). However, it is important to note that as dataset size increases, the relative advantages of these improvements may diminish (see Fig. \ref{fig:data_ablation}).

\begin{figure}[H]
  \centering
  % 第一行
  \begin{subfigure}[t]{0.49\textwidth}
    \vtop{\null\hbox{\includegraphics[width=\textwidth]{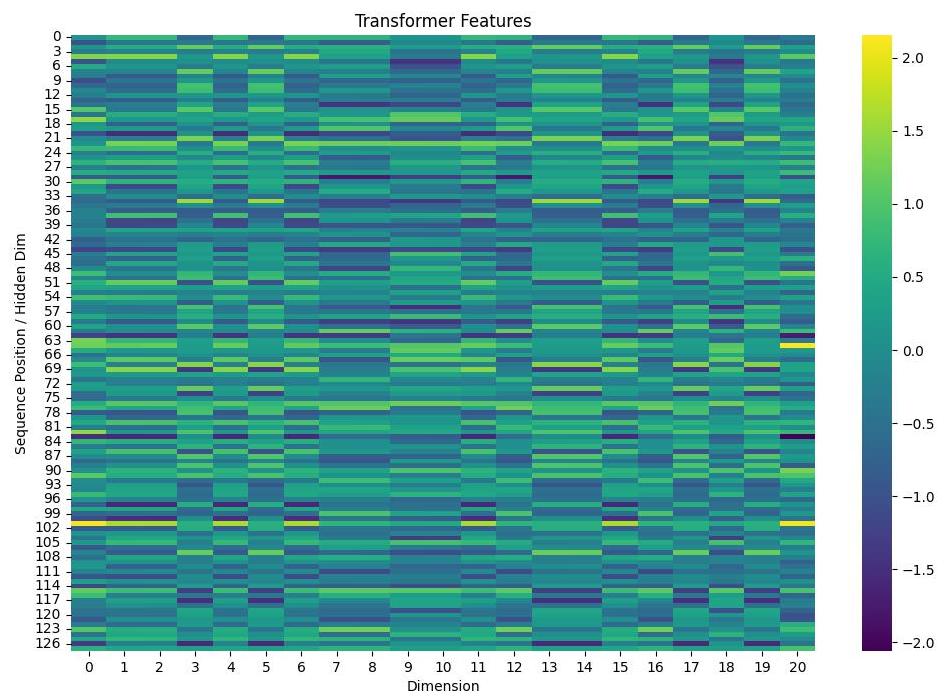}}}
    \caption{}
    \label{fig:transformer_feature}
  \end{subfigure}
   % \hfill % 添加空白或者 \hspace{5mm} 来调整子图间的水平间距
  \begin{subfigure}[t]{0.49\textwidth}
    \vtop{\null\hbox{\includegraphics[width=\textwidth]{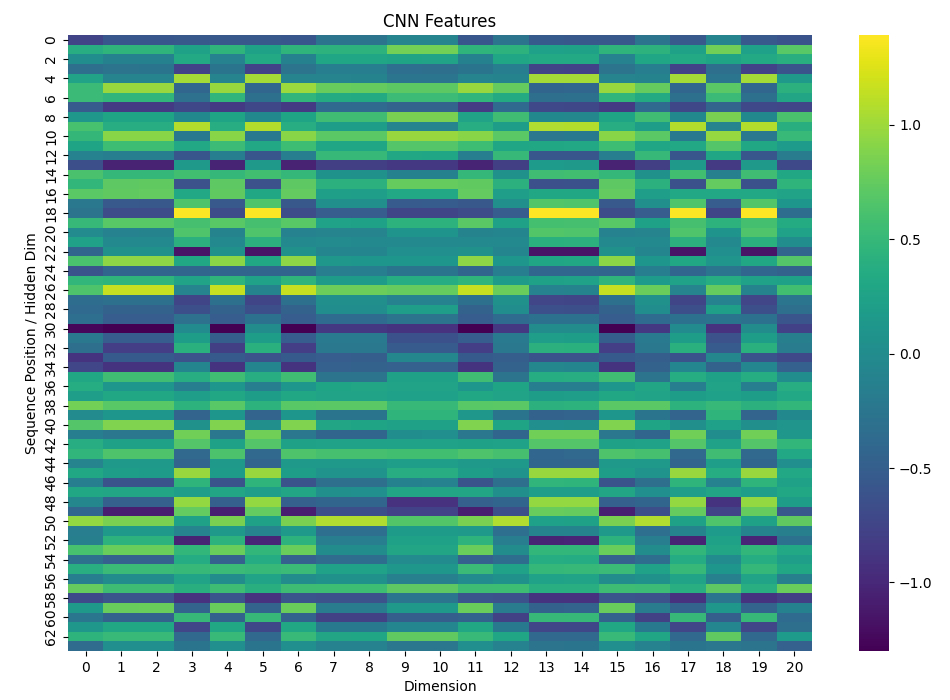}}}
    \caption{}
    \label{fig:cnn_feature}
  \end{subfigure}
 
\caption{Feature Maps. (a) The Transformer module is primarily responsible for capturing global features of siRNA sequences.The feature map illustrates the learned representations across various sequence positions and dimensions. For instance, the brighter yellow regions at specific positions and dimensions imply that the Transformer has recognized these as more significant features within the global context of the siRNA sequence. (b) The ConvNeXt module focuses on extracting local features from the siRNA sequences. The feature map obtained from the ConvNeXt module reveals distinct patterns. The distinct patterns and variations in color intensity highlight the local feature extraction capabilities of ConvNeXt. For example, certain regions with concentrated yellow patches might indicate the presence of specific local motifs that are highly relevant for predicting siRNA knockdown efficiency. }

  \label{fig:feature}
\end{figure}

Our research also highlights that focusing solely on the siRNA sequence allows the model to achieve excellent results, showcasing strong generalization capabilities. Additionally, thoughtful model design can further elevate performance (see Fig. \ref{fig:model_ablation}). Specifically, DeepSilencer integrates Transformer and ConvNeXt modules to leverage their strengths for enhanced prediction. The Transformer module captures global features of siRNA sequences via its self-attention mechanism, understanding long-range dependencies and interactions across the entire sequence(see Fig. \ref{fig:transformer_feature}). In contrast, the ConvNeXt module focuses on extracting local features, detecting specific short subsequences or local structural motifs within the siRNA(see Fig. \ref{fig:cnn_feature}). The combination of these two modules creates a powerful synergy, enabling the model to comprehensively understand siRNA sequences from both overall and local perspectives, which is a key factor in its superior performance.

\begin{figure}[H]
  \centering
  \includegraphics[width=0.8\textwidth]{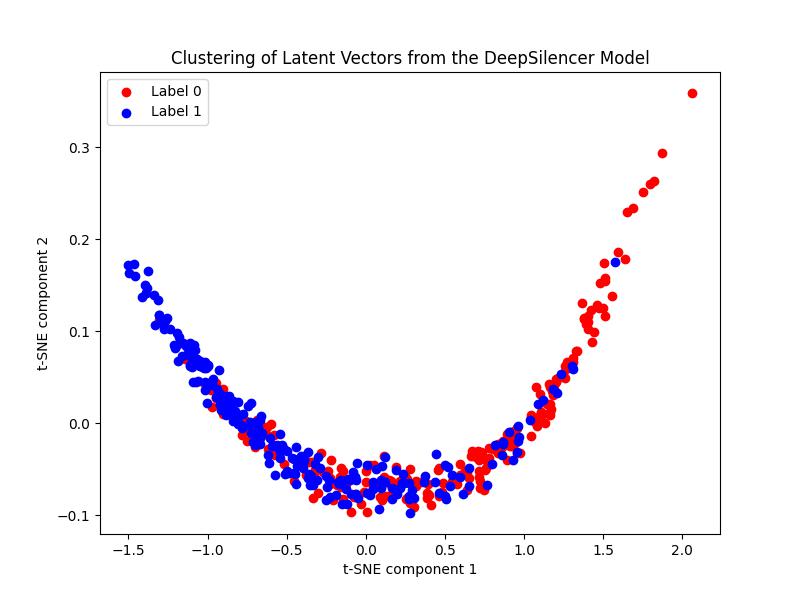}
  \caption{Clustering of latent vectors from the DeepSilencer Model. The x-axis represents the first t-SNE component, and the y-axis represents the second t-SNE component. The data points, colored based on their labels (red for Label 0 and blue for Label 1 corresponding to different knockdown efficiencies), form distinct clusters.}
  \label{fig:clusterling}
\end{figure}

To further visualize how the model separates siRNAs, we employed the t-SNE dimensionality reduction technique. The distribution of the points in the t-SNE plot reveals several interesting patterns(see Fig. \ref{fig:clusterling}). We observe that the points are not randomly scattered but rather form distinct clusters and gradients. The presence of clusters indicates that the DeepSilencer model has learned to group siRNAs with similar characteristics and can differentiate between those with different knockdown efficiencies to some extent. Although there is some overlap between the two classes, the overall trend shows that the model has learned to map siRNAs with higher knockdown efficiency differently from those with lower knockdown efficiency  in the reduced-dimensional space. This visualization validates the model's ability to learn discriminative representations and offers an intuitive understanding of its separation capabilities.

Examining the actual prediction results, as shown in Table~\ref{table:case_study}, we tested the model on the Mixset, which it had not encountered before. Initially, we were concerned that the use of relative position encoding in the transformer and CNN might cause the model to struggle with differentiating between similar sequences, especially those that are shifted versions of each other. However, DeepSilencer effectively distinguished between sequences with high similarity but significantly different knockdown efficiencies.

\begin{figure}[H]
  \centering
  \includegraphics[width=0.8\textwidth]{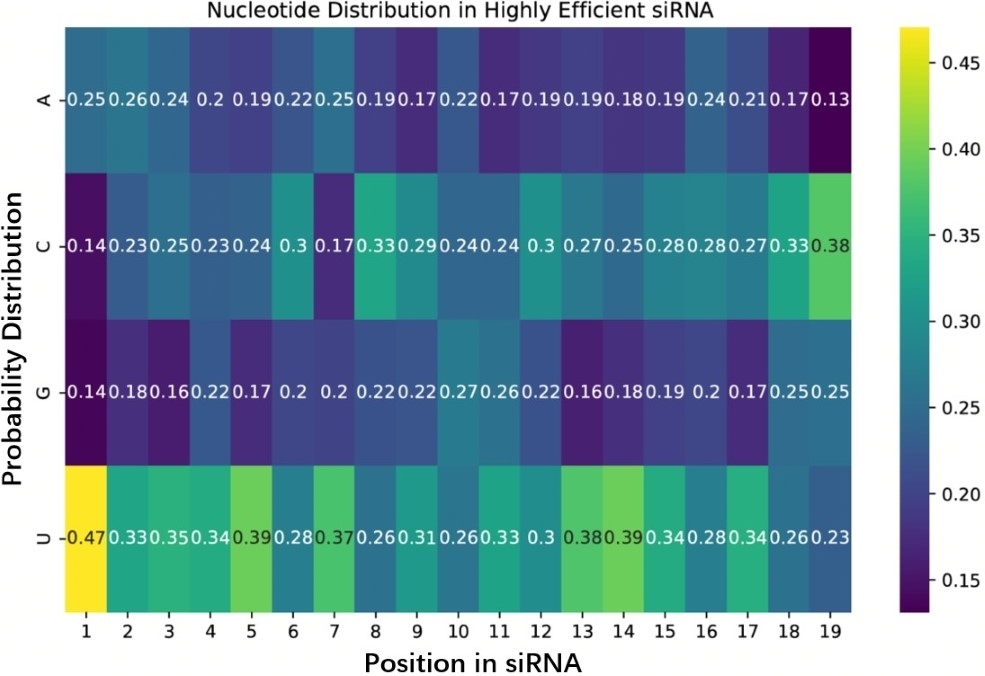}
  \caption{Nucleotide Distribution at Different Positions for High Knockdown Efficiency siRNAs in Mixset. The chart illustrates the likelihood of each nucleotide (A, U, G, C) appearing at specific positions within siRNAs that exhibit high knockdown efficiency. Notably, the first nucleotide is more frequently U, while G and C are less common. Conversely, the last position shows a higher occurrence of C and a lower occurrence of A, highlighting distinct sequence preferences associated with effective siRNA activity.}
  \label{fig:rna_dist}
\end{figure}

Focusing on specific samples, we observed that siRNAs in Mixset with high knockdown efficiency are more likely to start with a U and less likely with a G or C, while the last nucleotide is more likely to be a C and less likely an A (see Fig \ref{fig:rna_dist}). We identified two examples that contradict these statistics: one starting with G and ending with A but with high efficiency, and another starting with U and ending with C but with low efficiency. As expected, the model's performance on these cases was poor.

From analyzing these examples, we conclude that DeepSilencer has learned general sequence patterns related to siRNA knockdown efficiency from the available data. However, due to data limitations and our evolving understanding of the complex siRNA knockdown process, the model struggles with "anomalous" siRNAs.

In summary, DeepSilencer proves to be a robust and effective tool for siRNA knockdown prediction. The strategic use of Selective Pair Sampling and contrastive loss, along with a carefully crafted multi-task learning framework and the integration of Transformer and ConvNeXt modules, enables the model to excel, particularly with limited data. These insights contribute valuable advancements to RNA interference research, offering a promising direction for future siRNA prediction models.

\begin{table}[h]
\caption{Selected examples of knockout efficiency prediction on Mixset using the DeepSilencer}\label{table:case_study}
\begin{tabular*}{\textwidth}{@{\extracolsep\fill}lccc}
\toprule%
\multicolumn{4}{c}{Similar sequence comparison}  \\\cmidrule{1-4}%
siRNA & Regression prediction & Classification prediction & Experimental  \\
\midrule
UAAGCUUUCAUGGCAUCUU    & 0.6032  & 0.8523 & 0.6344\\
GUAAGCUUUCAUGGCAUCU    & 0.5022 & 0.4038  & 0.1508\\
UAUAGGAGGACCGUGUAGG    & 0.6394  & 0.8066 & 0.7283\\
GUAUAGGAGGACCGUGUAG    & 0.4446  & 0.4019 & 0.3961\\
UACUUAAUCAGAGACUUCA    & 0.7819  & 0.8007 & 0.9668\\
GUACUUAAUCAGAGACUUC    & 0.5336  & 0.5503 & 0.5046\\
\toprule%
\multicolumn{4}{c}{Examples of inaccurate predictions}  \\\cmidrule{1-4}%
siRNA & Regression prediction & Classification prediction & Experimental  \\
\midrule
GGCGUUGGUCGCUUCCGGA    & 0.4096  & 0.1418 & 0.9491\\
UUCUCUGGAAUGCCUGCAC    & 0.5395  & 0.8496 & 0.25\\
\botrule
\end{tabular*}
\end{table}

\section{Conclusion}\label{sec5}

We introduced DeepSilencer, a model for predicting siRNA knockdown efficiency, which demonstrates outstanding performance across multiple datasets, highlighting its robustness and adaptability. This success is attributed to techniques such as Selective Pair Sampling and contrastive loss within a multi-task learning framework.

As larger and more diverse datasets become available, we anticipate that siRNA prediction models will achieve even greater accuracy, which is crucial for developing precise siRNA-based therapies. Moreover, DeepSilencer's predictive capabilities could significantly aid drug development by optimizing therapeutic candidates and accelerating RNA interference-based treatments. Interestingly, some siRNAs that do not conform to typical high knockdown patterns also exhibit excellent efficiency. We hope future data and research will further enhance our model's capabilities.

Notably, while the primary focus of this study has been on siRNA, DeepSilencer holds great potential for other RNA-based therapeutics. MicroRNAs (miRNAs) regulate numerous biological processes, short hairpin RNAs (shRNAs) are crucial for RNAi-based strategies, and PIWI-interacting RNAs (piRNAs) are important in  transposon silencing \cite{ambros2004functions},\cite{jy2002rna},\cite{aravin2006novel},\cite{m2018deepmirtar} and \cite{khan2022deeppirna}. Given the sequence-based functional similarities among siRNAs, miRNAs, shRNAs, and piRNAs, DeepSilencer's architecture, which can capture sequence-related features for knockdown prediction, could be adapted to predict miRNA-mediated gene regulation, shRNA-mediated gene silencing and piRNA-related functions like target interactions. This may offer insights into miRNA-related diseases, facilitate miRNA-based therapies like designing better mimics or inhibitors, accelerate shRNA-based gene therapy development for genetic disorders and cancers, and open new research paths for piRNA-associated biology and potential therapies in germline-related diseases.

In summary, DeepSilencer marks a significant advancement in siRNA prediction and opens new avenues for future research and therapeutic advancements in RNA interference. It can accurately rank candidate siRNA sequences using only the sequences themselves and achieves even greater accuracy by incorporating simple thermodynamic parameters. This capability enhances drug development efficiency and reduces costs. Additionally, its potential applications in miRNA and shRNA research broaden its scope and relevance, offering exciting possibilities for further exploration in the field of RNA therapeutics.

\backmatter

\section*{Declarations}

\subsection*{Ethics approval and consent to participate}
Not applicable.

\subsection*{Consent for publication}
Not applicable.

\subsection*{Availability of data and materials}
The datasets supporting the conclusions of this article are available in \url{https://github.com/BlackCattt9/DeepSilencer/tree/main/data}. The datasets used in this article are sourced from the compiled datasets of oligoformer\cite{bai2024oligoformer}, which you can access from \url{https://github.com/lulab/OligoFormer}. These datasets compile nine datasets from the studies by Huesken\cite{huesken2005design}, Takayuki\cite{katoh2007specific}, Amarzguioui\cite{amarzguioui2003tolerance}, Haborth\cite{harborth2003sequence},  Khvorova, Reynolds\cite{reynolds2004rational}, Vickers\cite{vickers2003efficient},UiTei\cite{ui2004guidelines} and Hsieh\cite{hsieh2004library}.

The code used in this article are available on GitHub and can be accessed at the following link: \url{https://github.com/BlackCattt9/DeepSilencer} (we will organize and upload the code).

\subsection*{Competing interests}
The authors declare no competing interests

\subsection*{Funding}
This research is supported by the Research and Development Program of Ministry of Science and Technology of China (ID: 2020AAA0105800).

\subsection*{Authors' contributions}
WL was primarily responsible for data processing, model construction, coding, computation, and manuscript drafting. WW contributed significantly to the conceptualization of the research idea, analysis and discussion of the results, and manuscript revision. All authors read and approved the final manuscript.

\subsection*{Acknowledgements}
Not applicable.

\bibliography{sn-bibliography}% common bib file


\begin{thebibliography}{36}
\ifx \bisbn   \undefined \def \bisbn  #1{ISBN #1}\fi
\ifx \binits  \undefined \def \binits#1{#1}\fi
\ifx \bauthor  \undefined \def \bauthor#1{#1}\fi
\ifx \batitle  \undefined \def \batitle#1{#1}\fi
\ifx \bjtitle  \undefined \def \bjtitle#1{#1}\fi
\ifx \bvolume  \undefined \def \bvolume#1{\textbf{#1}}\fi
\ifx \byear  \undefined \def \byear#1{#1}\fi
\ifx \bissue  \undefined \def \bissue#1{#1}\fi
\ifx \bfpage  \undefined \def \bfpage#1{#1}\fi
\ifx \blpage  \undefined \def \blpage #1{#1}\fi
\ifx \burl  \undefined \def \burl#1{\textsf{#1}}\fi
\ifx \doiurl  \undefined \def \doiurl#1{\url{https://doi.org/#1}}\fi
\ifx \betal  \undefined \def \betal{\textit{et al.}}\fi
\ifx \binstitute  \undefined \def \binstitute#1{#1}\fi
\ifx \binstitutionaled  \undefined \def \binstitutionaled#1{#1}\fi
\ifx \bctitle  \undefined \def \bctitle#1{#1}\fi
\ifx \beditor  \undefined \def \beditor#1{#1}\fi
\ifx \bpublisher  \undefined \def \bpublisher#1{#1}\fi
\ifx \bbtitle  \undefined \def \bbtitle#1{#1}\fi
\ifx \bedition  \undefined \def \bedition#1{#1}\fi
\ifx \bseriesno  \undefined \def \bseriesno#1{#1}\fi
\ifx \blocation  \undefined \def \blocation#1{#1}\fi
\ifx \bsertitle  \undefined \def \bsertitle#1{#1}\fi
\ifx \bsnm \undefined \def \bsnm#1{#1}\fi
\ifx \bsuffix \undefined \def \bsuffix#1{#1}\fi
\ifx \bparticle \undefined \def \bparticle#1{#1}\fi
\ifx \barticle \undefined \def \barticle#1{#1}\fi
\bibcommenthead
\ifx \bconfdate \undefined \def \bconfdate #1{#1}\fi
\ifx \botherref \undefined \def \botherref #1{#1}\fi
\ifx \url \undefined \def \url#1{\textsf{#1}}\fi
\ifx \bchapter \undefined \def \bchapter#1{#1}\fi
\ifx \bbook \undefined \def \bbook#1{#1}\fi
\ifx \bcomment \undefined \def \bcomment#1{#1}\fi
\ifx \oauthor \undefined \def \oauthor#1{#1}\fi
\ifx \citeauthoryear \undefined \def \citeauthoryear#1{#1}\fi
\ifx \endbibitem  \undefined \def \endbibitem {}\fi
\ifx \bconflocation  \undefined \def \bconflocation#1{#1}\fi
\ifx \arxivurl  \undefined \def \arxivurl#1{\textsf{#1}}\fi
\csname PreBibitemsHook\endcsname

%%% 1
\bibitem[\protect\citeauthoryear{Zamore et~al.}{2000}]{zamore2000rnai}
\begin{barticle}
\bauthor{\bsnm{Zamore}, \binits{P.D.}},
\bauthor{\bsnm{Tuschl}, \binits{T.}},
\bauthor{\bsnm{Sharp}, \binits{P.A.}},
\bauthor{\bsnm{Bartel}, \binits{D.P.}}:
\batitle{Rnai: double-stranded rna directs the atp-dependent cleavage of mrna at 21 to 23 nucleotide intervals}.
\bjtitle{cell}
\bvolume{101}(\bissue{1}),
\bfpage{25}--\blpage{33}
(\byear{2000})
\doiurl{10.1016/S0092-8674(00)80620-0}
\end{barticle}
\endbibitem

%%% 2
\bibitem[\protect\citeauthoryear{Tang and Khvorova}{2024}]{tang2024rnai}
\begin{barticle}
\bauthor{\bsnm{Tang}, \binits{Q.}},
\bauthor{\bsnm{Khvorova}, \binits{A.}}:
\batitle{Rnai-based drug design: considerations and future directions}.
\bjtitle{Nature Reviews Drug Discovery}
\bvolume{23}(\bissue{5}),
\bfpage{341}--\blpage{364}
(\byear{2024})
\doiurl{10.1038/s41573-024-00912-9}
\end{barticle}
\endbibitem

%%% 3
\bibitem[\protect\citeauthoryear{Pratt and MacRae}{2009}]{pratt2009rna}
\begin{barticle}
\bauthor{\bsnm{Pratt}, \binits{A.J.}},
\bauthor{\bsnm{MacRae}, \binits{I.J.}}:
\batitle{The rna-induced silencing complex: a versatile gene-silencing machine}.
\bjtitle{Journal of Biological Chemistry}
\bvolume{284}(\bissue{27}),
\bfpage{17897}--\blpage{17901}
(\byear{2009})
\doiurl{10.1074/jbc.R900012200}
\end{barticle}
\endbibitem

%%% 4
\bibitem[\protect\citeauthoryear{Deerberg et~al.}{2013}]{deerberg2013minimal}
\begin{barticle}
\bauthor{\bsnm{Deerberg}, \binits{A.}},
\bauthor{\bsnm{Willkomm}, \binits{S.}},
\bauthor{\bsnm{Restle}, \binits{T.}}:
\batitle{Minimal mechanistic model of sirna-dependent target rna slicing by recombinant human argonaute 2 protein}.
\bjtitle{Proceedings of the National Academy of Sciences}
\bvolume{110}(\bissue{44}),
\bfpage{17850}--\blpage{17855}
(\byear{2013})
\doiurl{10.1073/pnas.121783811}
\end{barticle}
\endbibitem

%%% 5
\bibitem[\protect\citeauthoryear{Lu and Mathews}{2008}]{lu2008oligowalk}
\begin{barticle}
\bauthor{\bsnm{Lu}, \binits{Z.J.}},
\bauthor{\bsnm{Mathews}, \binits{D.H.}}:
\batitle{Oligowalk: an online sirna design tool utilizing hybridization thermodynamics}.
\bjtitle{Nucleic acids research}
\bvolume{36}(\bissue{suppl\_2}),
\bfpage{104}--\blpage{108}
(\byear{2008})
\doiurl{10.1093/nar/gkn250}
\end{barticle}
\endbibitem

%%% 6
\bibitem[\protect\citeauthoryear{Han et~al.}{2017}]{han2017utilizing}
\begin{barticle}
\bauthor{\bsnm{Han}, \binits{Y.}},
\bauthor{\bsnm{Liu}, \binits{Y.}},
\bauthor{\bsnm{Zhang}, \binits{H.}},
\bauthor{\bsnm{He}, \binits{F.}},
\bauthor{\bsnm{Shu}, \binits{C.}},
\bauthor{\bsnm{Dong}, \binits{L.}}:
\batitle{Utilizing selected di-and trinucleotides of sirna to predict rnai activity}.
\bjtitle{Computational and Mathematical Methods in Medicine}
\bvolume{2017}(\bissue{1}),
\bfpage{5043984}
(\byear{2017})
\doiurl{10.1155/2017/5043984}
\end{barticle}
\endbibitem

%%% 7
\bibitem[\protect\citeauthoryear{Ichihara et~al.}{2007}]{ichihara2007thermodynamic}
\begin{barticle}
\bauthor{\bsnm{Ichihara}, \binits{M.}},
\bauthor{\bsnm{Murakumo}, \binits{Y.}},
\bauthor{\bsnm{Masuda}, \binits{A.}},
\bauthor{\bsnm{Matsuura}, \binits{T.}},
\bauthor{\bsnm{Asai}, \binits{N.}},
\bauthor{\bsnm{Jijiwa}, \binits{M.}},
\bauthor{\bsnm{Ishida}, \binits{M.}},
\bauthor{\bsnm{Shinmi}, \binits{J.}},
\bauthor{\bsnm{Yatsuya}, \binits{H.}},
\bauthor{\bsnm{Qiao}, \binits{S.}}, \betal:
\batitle{Thermodynamic instability of sirna duplex is a prerequisite for dependable prediction of sirna activities}.
\bjtitle{Nucleic acids research}
\bvolume{35}(\bissue{18}),
\bfpage{123}
(\byear{2007})
\doiurl{10.1093/nar/gkm699}
\end{barticle}
\endbibitem

%%% 8
\bibitem[\protect\citeauthoryear{Lu and Mathews}{2008a}]{lu2008efficient}
\begin{barticle}
\bauthor{\bsnm{Lu}, \binits{Z.J.}},
\bauthor{\bsnm{Mathews}, \binits{D.H.}}:
\batitle{Efficient sirna selection using hybridization thermodynamics}.
\bjtitle{Nucleic acids research}
\bvolume{36}(\bissue{2}),
\bfpage{640}--\blpage{647}
(\byear{2008})
\doiurl{10.1093/nar/gkm920}
\end{barticle}
\endbibitem

%%% 9
\bibitem[\protect\citeauthoryear{Lu and Mathews}{2008b}]{lu2008fundamental}
\begin{barticle}
\bauthor{\bsnm{Lu}, \binits{Z.J.}},
\bauthor{\bsnm{Mathews}, \binits{D.H.}}:
\batitle{Fundamental differences in the equilibrium considerations for sirna and antisense oligodeoxynucleotide design}.
\bjtitle{Nucleic Acids Research}
\bvolume{36}(\bissue{11}),
\bfpage{3738}--\blpage{3745}
(\byear{2008})
\doiurl{10.1093/nar/gkn266}
\end{barticle}
\endbibitem

%%% 10
\bibitem[\protect\citeauthoryear{Han et~al.}{2018}]{han2018sirna}
\begin{barticle}
\bauthor{\bsnm{Han}, \binits{Y.}},
\bauthor{\bsnm{He}, \binits{F.}},
\bauthor{\bsnm{Chen}, \binits{Y.}},
\bauthor{\bsnm{Liu}, \binits{Y.}},
\bauthor{\bsnm{Yu}, \binits{H.}}:
\batitle{Sirna silencing efficacy prediction based on a deep architecture}.
\bjtitle{BMC genomics}
\bvolume{19},
\bfpage{59}--\blpage{65}
(\byear{2018})
\doiurl{10.1186/s12864-018-5028-8}
\end{barticle}
\endbibitem

%%% 11
\bibitem[\protect\citeauthoryear{La~Rosa et~al.}{2022}]{ijms232214211}
\begin{botherref}
\oauthor{\bsnm{La~Rosa}, \binits{M.}},
\oauthor{\bsnm{Fiannaca}, \binits{A.}},
\oauthor{\bsnm{La~Paglia}, \binits{L.}},
\oauthor{\bsnm{Urso}, \binits{A.}}:
A graph neural network approach for the analysis of sirna-target biological networks.
International Journal of Molecular Sciences
\textbf{23}(22)
(2022)
\doiurl{10.3390/ijms232214211}
\end{botherref}
\endbibitem

%%% 12
\bibitem[\protect\citeauthoryear{Vaswani et~al.}{2017}]{10.5555/3295222.3295349}
\begin{bchapter}
\bauthor{\bsnm{Vaswani}, \binits{A.}},
\bauthor{\bsnm{Shazeer}, \binits{N.}},
\bauthor{\bsnm{Parmar}, \binits{N.}},
\bauthor{\bsnm{Uszkoreit}, \binits{J.}},
\bauthor{\bsnm{Jones}, \binits{L.}},
\bauthor{\bsnm{Gomez}, \binits{A.N.}},
\bauthor{\bsnm{Kaiser}, \binits{L.}},
\bauthor{\bsnm{Polosukhin}, \binits{I.}}:
\bctitle{Attention is all you need}.
In: \bbtitle{Proceedings of the 31st International Conference on Neural Information Processing Systems}.
\bsertitle{NIPS'17},
pp. \bfpage{6000}--\blpage{6010}.
\bpublisher{Curran Associates Inc.},
\blocation{Red Hook, NY, USA}
(\byear{2017}).
\doiurl{10.5555/3295222.3295349}
\end{bchapter}
\endbibitem

%%% 13
\bibitem[\protect\citeauthoryear{Devlin}{2018}]{devlin2018bert}
\begin{barticle}
\bauthor{\bsnm{Devlin}, \binits{J.}}:
\batitle{Bert: Pre-training of deep bidirectional transformers for language understanding}.
\bjtitle{arXiv preprint arXiv:1810.04805}
(\byear{2018})
\doiurl{10.18653/v1/N19-1423}
\end{barticle}
\endbibitem

%%% 14
\bibitem[\protect\citeauthoryear{Xu et~al.}{2024}]{xu2024bert}
\begin{barticle}
\bauthor{\bsnm{Xu}, \binits{J.}},
\bauthor{\bsnm{Xu}, \binits{N.}},
\bauthor{\bsnm{Xie}, \binits{W.}},
\bauthor{\bsnm{Zhao}, \binits{C.}},
\bauthor{\bsnm{Yu}, \binits{L.}},
\bauthor{\bsnm{Feng}, \binits{W.}}:
\batitle{Bert-sirna: sirna target prediction based on bert pre-trained interpretable model}.
\bjtitle{Gene}
\bvolume{910},
\bfpage{148330}
(\byear{2024})
\doiurl{10.1016/j.gene.2024.148330}
\end{barticle}
\endbibitem

%%% 15
\bibitem[\protect\citeauthoryear{Hochreiter}{1997}]{hochreiter1997long}
\begin{botherref}
\oauthor{\bsnm{Hochreiter}, \binits{S.}}:
Long short-term memory.
Neural Computation MIT-Press
(1997)
\end{botherref}
\endbibitem

%%% 16
\bibitem[\protect\citeauthoryear{LeCun et~al.}{1989}]{lecun1989backpropagation}
\begin{barticle}
\bauthor{\bsnm{LeCun}, \binits{Y.}},
\bauthor{\bsnm{Boser}, \binits{B.}},
\bauthor{\bsnm{Denker}, \binits{J.S.}},
\bauthor{\bsnm{Henderson}, \binits{D.}},
\bauthor{\bsnm{Howard}, \binits{R.E.}},
\bauthor{\bsnm{Hubbard}, \binits{W.}},
\bauthor{\bsnm{Jackel}, \binits{L.D.}}:
\batitle{Backpropagation applied to handwritten zip code recognition}.
\bjtitle{Neural computation}
\bvolume{1}(\bissue{4}),
\bfpage{541}--\blpage{551}
(\byear{1989})
\doiurl{10.1162/neco.1989.1.4.541}
\end{barticle}
\endbibitem

%%% 17
\bibitem[\protect\citeauthoryear{Bai et~al.}{2024}]{bai2024oligoformer}
\begin{botherref}
\oauthor{\bsnm{Bai}, \binits{Y.}},
\oauthor{\bsnm{Zhong}, \binits{H.}},
\oauthor{\bsnm{Wang}, \binits{T.}},
\oauthor{\bsnm{Lu}, \binits{Z.J.}}:
Oligoformer: an accurate and robust prediction method for sirna design.
bioRxiv,
2024--02
(2024)
\doiurl{10.1101/2024.02.02.578533}
\end{botherref}
\endbibitem

%%% 18
\bibitem[\protect\citeauthoryear{Huesken et~al.}{2005}]{huesken2005design}
\begin{barticle}
\bauthor{\bsnm{Huesken}, \binits{D.}},
\bauthor{\bsnm{Lange}, \binits{J.}},
\bauthor{\bsnm{Mickanin}, \binits{C.}},
\bauthor{\bsnm{Weiler}, \binits{J.}},
\bauthor{\bsnm{Asselbergs}, \binits{F.}},
\bauthor{\bsnm{Warner}, \binits{J.}},
\bauthor{\bsnm{Meloon}, \binits{B.}},
\bauthor{\bsnm{Engel}, \binits{S.}},
\bauthor{\bsnm{Rosenberg}, \binits{A.}},
\bauthor{\bsnm{Cohen}, \binits{D.}}, \betal:
\batitle{Design of a genome-wide sirna library using an artificial neural network}.
\bjtitle{Nature biotechnology}
\bvolume{23}(\bissue{8}),
\bfpage{995}--\blpage{1001}
(\byear{2005})
\doiurl{10.1038/nbt1118}
\end{barticle}
\endbibitem

%%% 19
\bibitem[\protect\citeauthoryear{Katoh and Suzuki}{2007}]{katoh2007specific}
\begin{barticle}
\bauthor{\bsnm{Katoh}, \binits{T.}},
\bauthor{\bsnm{Suzuki}, \binits{T.}}:
\batitle{Specific residues at every third position of sirna shape its efficient rnai activity}.
\bjtitle{Nucleic acids research}
\bvolume{35}(\bissue{4}),
\bfpage{27}
(\byear{2007})
\doiurl{10.1093/nar/gkl1120}
\end{barticle}
\endbibitem

%%% 20
\bibitem[\protect\citeauthoryear{Amarzguioui et~al.}{2003}]{amarzguioui2003tolerance}
\begin{barticle}
\bauthor{\bsnm{Amarzguioui}, \binits{M.}},
\bauthor{\bsnm{Holen}, \binits{T.}},
\bauthor{\bsnm{Babaie}, \binits{E.}},
\bauthor{\bsnm{Prydz}, \binits{H.}}:
\batitle{Tolerance for mutations and chemical modifications in a sirna}.
\bjtitle{Nucleic acids research}
\bvolume{31}(\bissue{2}),
\bfpage{589}--\blpage{595}
(\byear{2003})
\doiurl{10.1093/nar/gkg147}
\end{barticle}
\endbibitem

%%% 21
\bibitem[\protect\citeauthoryear{Harborth et~al.}{2003}]{harborth2003sequence}
\begin{barticle}
\bauthor{\bsnm{Harborth}, \binits{J.}},
\bauthor{\bsnm{Elbashir}, \binits{S.M.}},
\bauthor{\bsnm{Vandenburgh}, \binits{K.}},
\bauthor{\bsnm{Manninga}, \binits{H.}},
\bauthor{\bsnm{Scaringe}, \binits{S.A.}},
\bauthor{\bsnm{Weber}, \binits{K.}},
\bauthor{\bsnm{Tuschl}, \binits{T.}}:
\batitle{Sequence, chemical, and structural variation of small interfering rnas and short hairpin rnas and the effect on mammalian gene silencing}.
\bjtitle{Antisense and Nucleic Acid Drug Development}
\bvolume{13}(\bissue{2}),
\bfpage{83}--\blpage{105}
(\byear{2003})
\doiurl{10.1089/10872900332162963}
\end{barticle}
\endbibitem

%%% 22
\bibitem[\protect\citeauthoryear{Reynolds et~al.}{2004}]{reynolds2004rational}
\begin{barticle}
\bauthor{\bsnm{Reynolds}, \binits{A.}},
\bauthor{\bsnm{Leake}, \binits{D.}},
\bauthor{\bsnm{Boese}, \binits{Q.}},
\bauthor{\bsnm{Scaringe}, \binits{S.}},
\bauthor{\bsnm{Marshall}, \binits{W.S.}},
\bauthor{\bsnm{Khvorova}, \binits{A.}}:
\batitle{Rational sirna design for rna interference}.
\bjtitle{Nature biotechnology}
\bvolume{22}(\bissue{3}),
\bfpage{326}--\blpage{330}
(\byear{2004})
\doiurl{10.1038/nbt936}
\end{barticle}
\endbibitem

%%% 23
\bibitem[\protect\citeauthoryear{Vickers et~al.}{2003}]{vickers2003efficient}
\begin{barticle}
\bauthor{\bsnm{Vickers}, \binits{T.A.}},
\bauthor{\bsnm{Koo}, \binits{S.}},
\bauthor{\bsnm{Bennett}, \binits{C.F.}},
\bauthor{\bsnm{Crooke}, \binits{S.T.}},
\bauthor{\bsnm{Dean}, \binits{N.M.}},
\bauthor{\bsnm{Baker}, \binits{B.F.}}:
\batitle{Efficient reduction of target rnas by small interfering rna and rnase h-dependent antisense agents: a comparative analysis}.
\bjtitle{Journal of Biological Chemistry}
\bvolume{278}(\bissue{9}),
\bfpage{7108}--\blpage{7118}
(\byear{2003})
\doiurl{10.1074/jbc.M210326200}
\end{barticle}
\endbibitem

%%% 24
\bibitem[\protect\citeauthoryear{Ui-Tei et~al.}{2004}]{ui2004guidelines}
\begin{barticle}
\bauthor{\bsnm{Ui-Tei}, \binits{K.}},
\bauthor{\bsnm{Naito}, \binits{Y.}},
\bauthor{\bsnm{Takahashi}, \binits{F.}},
\bauthor{\bsnm{Haraguchi}, \binits{T.}},
\bauthor{\bsnm{Ohki-Hamazaki}, \binits{H.}},
\bauthor{\bsnm{Juni}, \binits{A.}},
\bauthor{\bsnm{Ueda}, \binits{R.}},
\bauthor{\bsnm{Saigo}, \binits{K.}}:
\batitle{Guidelines for the selection of highly effective sirna sequences for mammalian and chick rna interference}.
\bjtitle{Nucleic acids research}
\bvolume{32}(\bissue{3}),
\bfpage{936}--\blpage{948}
(\byear{2004})
\doiurl{10.1093/nar/gkh247}
\end{barticle}
\endbibitem

%%% 25
\bibitem[\protect\citeauthoryear{Hsieh et~al.}{2004}]{hsieh2004library}
\begin{barticle}
\bauthor{\bsnm{Hsieh}, \binits{A.C.}},
\bauthor{\bsnm{Bo}, \binits{R.}},
\bauthor{\bsnm{Manola}, \binits{J.}},
\bauthor{\bsnm{Vazquez}, \binits{F.}},
\bauthor{\bsnm{Bare}, \binits{O.}},
\bauthor{\bsnm{Khvorova}, \binits{A.}},
\bauthor{\bsnm{Scaringe}, \binits{S.}},
\bauthor{\bsnm{Sellers}, \binits{W.R.}}:
\batitle{A library of sirna duplexes targeting the phosphoinositide 3-kinase pathway: determinants of gene silencing for use in cell-based screens}.
\bjtitle{Nucleic acids research}
\bvolume{32}(\bissue{3}),
\bfpage{893}--\blpage{901}
(\byear{2004})
\doiurl{10.1093/nar/gkh238}
\end{barticle}
\endbibitem

%%% 26
\bibitem[\protect\citeauthoryear{Kihyuk.S}{2016}]{NIPS2016_6b180037}
\begin{bchapter}
\bauthor{\bsnm{Kihyuk.S}}:
\bctitle{Improved deep metric learning with multi-class n-pair loss objective}.
In: \bbtitle{Advances in Neural Information Processing Systems}.
\bsertitle{NIPS'16},
vol. \bseriesno{29},
pp. \bfpage{1857}--\blpage{1865}
(\byear{2016}).
\doiurl{10.5555/3157096.3157304}
\end{bchapter}
\endbibitem

%%% 27
\bibitem[\protect\citeauthoryear{Zhang and Yang}{2018}]{zhang2018overview}
\begin{barticle}
\bauthor{\bsnm{Zhang}, \binits{Y.}},
\bauthor{\bsnm{Yang}, \binits{Q.}}:
\batitle{An overview of multi-task learning}.
\bjtitle{National Science Review}
\bvolume{5}(\bissue{1}),
\bfpage{30}--\blpage{43}
(\byear{2018})
\doiurl{10.1093/nsr/nwx105}
\end{barticle}
\endbibitem

%%% 28
\bibitem[\protect\citeauthoryear{Ren et~al.}{2017}]{7485869}
\begin{barticle}
\bauthor{\bsnm{Ren}, \binits{S.}},
\bauthor{\bsnm{He}, \binits{K.}},
\bauthor{\bsnm{Girshick}, \binits{R.}},
\bauthor{\bsnm{Sun}, \binits{J.}}:
\batitle{Faster r-cnn: Towards real-time object detection with region proposal networks}.
\bjtitle{IEEE Transactions on Pattern Analysis and Machine Intelligence}
\bvolume{39}(\bissue{6}),
\bfpage{1137}--\blpage{1149}
(\byear{2017})
\doiurl{10.1109/TPAMI.2016.2577031}
\end{barticle}
\endbibitem

%%% 29
\bibitem[\protect\citeauthoryear{Woo et~al.}{2023}]{woo2023convnext}
\begin{bchapter}
\bauthor{\bsnm{Woo}, \binits{S.}},
\bauthor{\bsnm{Debnath}, \binits{S.}},
\bauthor{\bsnm{Hu}, \binits{R.}},
\bauthor{\bsnm{Chen}, \binits{X.}},
\bauthor{\bsnm{Liu}, \binits{Z.}},
\bauthor{\bsnm{Kweon}, \binits{I.S.}},
\bauthor{\bsnm{Xie}, \binits{S.}}:
\bctitle{Convnext v2: Co-designing and scaling convnets with masked autoencoders}.
In: \bbtitle{Proceedings of the IEEE/CVF Conference on Computer Vision and Pattern Recognition},
pp. \bfpage{16133}--\blpage{16142}
(\byear{2023}).
\doiurl{10.1109/CVPR52729.2023.01548}
\end{bchapter}
\endbibitem

%%% 30
\bibitem[\protect\citeauthoryear{Dai et~al.}{2021}]{dai2021coatnet}
\begin{barticle}
\bauthor{\bsnm{Dai}, \binits{Z.}},
\bauthor{\bsnm{Liu}, \binits{H.}},
\bauthor{\bsnm{Le}, \binits{Q.V.}},
\bauthor{\bsnm{Tan}, \binits{M.}}:
\batitle{Coatnet: Marrying convolution and attention for all data sizes}.
\bjtitle{Advances in neural information processing systems}
\bvolume{34},
\bfpage{3965}--\blpage{3977}
(\byear{2021})
\doiurl{10.5555/3540261.3540564}
\end{barticle}
\endbibitem

%%% 31
\bibitem[\protect\citeauthoryear{Dagli}{2023}]{Dagli2023AstroformerMD}
\begin{botherref}
\oauthor{\bsnm{Dagli}, \binits{R.}}:
Astroformer: More data might not be all you need for classification.
ArXiv
\textbf{abs/2304.05350}
(2023)
\doiurl{10.48550/arXiv.2304.05350}
\end{botherref}
\endbibitem

%%% 32
\bibitem[\protect\citeauthoryear{Ambros}{2004}]{ambros2004functions}
\begin{barticle}
\bauthor{\bsnm{Ambros}, \binits{V.}}:
\batitle{The functions of animal micrornas}.
\bjtitle{Nature}
\bvolume{431}(\bissue{7006}),
\bfpage{350}--\blpage{355}
(\byear{2004})
\doiurl{10.1038/nature02871}
\end{barticle}
\endbibitem

%%% 33
\bibitem[\protect\citeauthoryear{Jy et~al.}{2002}]{jy2002rna}
\begin{botherref}
\oauthor{\bsnm{Jy}, \binits{Y.}},
\oauthor{\bsnm{Sl}, \binits{D.}},
\oauthor{\bsnm{Dl}, \binits{T.}}:
Rna interference by expression of short-interfering rnas and hairpin rnas in mammalian cells.
Proceedings of the National Academy of Sciences of the United States of America
\textbf{99}(9)
(2002)
\doiurl{10.1073/pnas.092143499}
\end{botherref}
\endbibitem

%%% 34
\bibitem[\protect\citeauthoryear{Aravin et~al.}{2006}]{aravin2006novel}
\begin{barticle}
\bauthor{\bsnm{Aravin}, \binits{A.}},
\bauthor{\bsnm{Gaidatzis}, \binits{D.}},
\bauthor{\bsnm{Pfeffer}, \binits{S.}},
\bauthor{\bsnm{{Lagos-Quintana}}, \binits{M.}},
\bauthor{\bsnm{Landgraf}, \binits{P.}},
\bauthor{\bsnm{Iovino}, \binits{N.}},
\bauthor{\bsnm{Morris}, \binits{P.}},
\bauthor{\bsnm{Brownstein}, \binits{M.J.}},
\bauthor{\bsnm{{Kuramochi-Miyagawa}}, \binits{S.}},
\bauthor{\bsnm{Nakano}, \binits{T.}},
\bauthor{\bsnm{Chien}, \binits{M.}},
\bauthor{\bsnm{Russo}, \binits{J.J.}},
\bauthor{\bsnm{Ju}, \binits{J.}},
\bauthor{\bsnm{Sheridan}, \binits{R.}},
\bauthor{\bsnm{Sander}, \binits{C.}},
\bauthor{\bsnm{Zavolan}, \binits{M.}},
\bauthor{\bsnm{Tuschl}, \binits{T.}}:
\batitle{A novel class of small rnas bind to mili protein in mouse testes}.
\bjtitle{Nature}
\bvolume{442}(\bissue{7099}),
\bfpage{203}--\blpage{207}
(\byear{2006})
\doiurl{10.1038/nature04916}
\end{barticle}
\endbibitem

%%% 35
\bibitem[\protect\citeauthoryear{M et~al.}{2018}]{m2018deepmirtar}
\begin{botherref}
\oauthor{\bsnm{M}, \binits{W.}},
\oauthor{\bsnm{P}, \binits{C.}},
\oauthor{\bsnm{Z}, \binits{Z.}},
\oauthor{\bsnm{H}, \binits{L.}},
\oauthor{\bsnm{T}, \binits{L.}}:
Deepmirtar: A deep-learning approach for predicting human mirna targets.
Bioinformatics (Oxford, England)
\textbf{34}(22)
(2018)
\doiurl{10.1093/bioinformatics/bty424}
\end{botherref}
\endbibitem

%%% 36
\bibitem[\protect\citeauthoryear{Khan et~al.}{2022}]{khan2022deeppirna}
\begin{barticle}
\bauthor{\bsnm{Khan}, \binits{S.}},
\bauthor{\bsnm{Khan}, \binits{M.}},
\bauthor{\bsnm{Iqbal}, \binits{N.}},
\bauthor{\bsnm{Rahman}, \binits{M.A.A.}},
\bauthor{\bsnm{Karim}, \binits{M.K.A.}}:
\batitle{Deep-pirna: Bi-layered prediction model for piwi-interacting rna using discriminative features}.
\bjtitle{Computers, Materials \& Continua}
\bvolume{72}(\bissue{2}),
\bfpage{2243}--\blpage{2258}
(\byear{2022})
\doiurl{10.32604/cmc.2022.022901}
\end{barticle}
\endbibitem

\end{thebibliography}

\end{document}